%% file: fermi_grg4.tex
\documentclass[nofootinbib,showpacs]{revtex4-1}
\usepackage[dvips]{graphicx}
\usepackage{amsmath,mathenv,latexsym,setspace,bm,tensor}
\usepackage{natbib}

\input{macro.tex}

\begin{document}

\title{Extended Fermi coordinates}
\author{P.~Delva}
\email{Pacome.Delva@esa.int}
\affiliation{European Space Agency, The Advanced Concepts Team \\ Keplerlaan 1, 2201 AZ Noordwijk, The Netherlands}
\author{M.-C.~Angonin}
\email{m-c.angonin@obspm.fr}
\affiliation{SYRTE, Observatoire de Paris, CNRS, UPMC, 61 avenue de l'Observatoire, 75014 Paris, France}

\begin{abstract}
We extend the notion of Fermi coordinates to a generalized definition in which the highest orders are described by arbitrary functions. From this definition rises a formalism that naturally gives coordinate transformation formulae. Some examples are developed in order to discuss the physical meaning of Fermi coordinates.
\end{abstract}

\maketitle

\section{Introduction}
\label{intro}
The Einstein Equivalence Principle postulates that, in presence of a gravitational field, the physical laws of Special Relativity are valid in an infinitesimally small laboratory. It implies that, in the theory of General Relativity, one should be able to locally reproduce the inertial frame of special relativity. Indeed, one can find, for each event $P$ of spacetime, a local inertial frame where the components of the metric tensor verify $g_{\mu \nu , \alpha}(P) = 0$~(Greek indices run from 0 to 3)\footnote{In this article, we limit our discussion to the four dimensional spacetime, but the results can be readily generalized to higher dimensions.}. The coordinates of such a frame are called {\it Riemann coordinates}\footnote{Riemann introduced in 1854 the so-called {\it normal Riemann coordinates}~\cite{riemann54} for the case of a Riemannian space. It has been generalized later to a quasi-Riemannian space.}.

A specialization of the Riemann coordinates is done when the coordinate lines going through the event $P$ are taken as geodesics. Such coordinates are called {\it Riemann normal coordinates}. According to \citeauthor{misner73}~\citep[p.285]{misner73}, they have the advantage of displaying ``beautiful ties'' to the Riemann curvature tensor\footnote{We use a different convention than \citet{misner73} for the Riemann tensor: $R\indices{^\mu_{\alpha \nu \beta}} = \Gamma\indices{^\mu_{\alpha \nu ,\beta}} - \Gamma\indices{^\mu _{\alpha \beta ,\nu}} - \Gamma\indices{^\mu _{\nu \sigma}} \Gamma\indices{^\sigma _{\alpha \beta}} + \Gamma\indices{^\mu _{\beta \sigma}} \Gamma\indices{^\sigma _{\alpha \nu}}$.}:
\be \Gamma\indices{^\alpha _{\beta \mu, \nu}} (P) = \dfrac{1}{3} \left( R\indices{^\alpha _{\beta \mu \nu}} + R\indices{^\alpha _{\mu \beta \nu}} \right) \ ; \ g_{\alpha \beta , \mu \nu} (P) = \dfrac{1}{3} \left( R_{\alpha \mu \beta \nu} + R_{\alpha \nu \beta \mu} \right) \ee
where $\Gamma\indices{^\alpha _{\beta \mu}}$ are the connection coefficients. Of course, there are many more inertial coordinate system than the Riemann normal one at event $P$. Any kind of new coordinate system is \emph{a priori} acceptable if the difference with the Riemann normal one concerns terms of order superior or equal to three. Choosing one among them is a personal choice, but can have consequences on the conception and the construction of an experiment. Indeed, to modelize an \emph{apparatus} one can define its constraints in a locally inertial coordinate system; but the form of the constraints will be different in different locally inertial coordinate systems\footnote{The principle of covariance does not imply that coordinate systems are not important in general relativity. It just helps to separate the difficult task of the conception and construction of an experiment, from the conception and construction of the coordinate system (see the discussion of \citet{coll06a}). One amazing fact is the lack of prescription for physical realizations of coordinate systems in general relativity, nearly one hundred years after its birth.}. It is important to know how the freedom in the choice of the locally inertial coordinate system changes the form of the constraints. In this paper, we investigate a similar problem when dealing with the Fermi coordinates.

The Fermi coordinates have been introduced by \citeauthor{fermi22a} in 1922~\citep{fermi22a,fermi22b,fermi22c}. They are defined in the neighborhood of the worldline $\C$ of an \emph{apparatus} or an observer. If the worldline is a geodesic, they display the same property as the Riemann coordinates all along the worldline $C$ (not only at a precise event): $g_{\mu \nu , \alpha}(\C) \equiv  \bar{g}_{\mu \nu , \alpha} = 0$ (in the following, the bar stands for the value of a function along a worldline $\C$). Another very useful consequence is that the time coordinate along $\C$ is the proper time of the observer. The Fermi frame, ie. the coordinate basis linked to the Fermi coordinates, is Fermi-Walker transported along $\C$. This frame is locally non rotating with respect to gyroscopes.

In the neighborhood of an arbitrary curve in spacetime, \citet{levicivita27a} showed that one can find local coordinates where $\bar{g}_{\mu \nu , \alpha} = 0$. This demonstration has often been attributed to \citeauthor{fermi22a}, while he only considered the case of a geodesic (see the detailed discussion of \citet{bini02}). However, the basis has to be parallel transported along the curve and the time coordinate does not coincide anymore with the proper time along the curve. \citet{oraifeartaigh58} extended the problem to submanifolds of dimension superior to one. It appears that Fermi coordinates only exist under very constrained conditions. For example, there is no simple submanifold of dimension superior to one in a Schwarzschild spacetime on which the connection coefficients and thus the $g_{\mu \nu , \alpha}$ vanish for given coordinates.

For the sake of simplicity, \citet{manasse63} introduced, along a geodesic, a specialization of the Fermi coordinates: the {\it Fermi normal coordinates}. The idea is similar to the Riemann normal coordinates; for each event $P$ on the worldline $\C$, the spatial coordinate lines crossing through $P$ are considered as geodesics. Then the second derivatives of the metric tensor components in the Fermi normal coordinates can be written as a combination of the curvature tensor components in the initial coordinates. Later \citet{ni78} extended this idea to an arbitrary worldline, adding terms depending on the acceleration and the angular velocity of the observer.

Nowadays the Fermi normal coordinates are usually - although improperly - called Fermi coordinates. In experimental gravitation, Fermi normal coordinates are a powerful tool used to describe various experiments: since the Fermi normal coordinates are Minkowskian to first order, the equations of physics in a Fermi normal frame are the ones of special relativity, plus corrections of higher order in the Fermi normal coordinates, therefore accounting for the gravitational field and its coupling to the inertial effects. Additionally, for small velocities $v$ compared to light velocity $c$, the Fermi normal coordinates can be assimilated to the zeroth order in $(v/c)$ to classical Galilean coordinates. They can be used to describe an \emph{apparatus} in a ``Newtonian'' way (e.g.~\citep{delva06,angonin06,ashby75,collas07}), or to interpret the outcome of an experiment (e.g.~\citep{diazmiguel04} and comment~\citep{linet04}, \citep{chicone05,chicone06b,kojima06,ishii05}). In these approaches, the Fermi normal coordinates are considered to have a physical meaning, coming from the principle of equivalence (see e.g.~\citep{leaute83}), and an operational meaning: the Fermi normal frame can be realized with an ideal clock and a non extensible thread~\citep{tourrenc99a}. This justifies the fact that they are used to define an \emph{apparatus} or the result of an experiment in terms of coordinate dependent quantities. However, the form of the metric corrections of higher order in a Fermi normal frame does not depend on some physical assumptions but on mathematical considerations (mainly simplicity). It is therefore legitimate to widen the definition of the Fermi normal coordinates in order to discuss their physical meaning.

As for Riemann normal coordinates, the Fermi normal coordinates are not the only ones reproducing locally (along $\C$) the frame of special relativity. The goal of this paper is to extend the usual definition - given by \citet{manasse63} - of the Fermi coordinates. This leads to what we will call in the sequel the \emph{extended Fermi coordinates}. As in \citep{delva07} and \citep{klein08}, we start from Taylor expansions of the coordinate transformations. This approach is more ``cumbersome'' than the method of \citeauthor{manasse63}, but it has the advantage to give the link between the initial coordinates to the extended Fermi coordinates. In section~\ref{sec:1}, we fix the form of the coordinate transformations up to the second order in the extended Fermi coordinates, only relying on physical assumptions coming from the principle of equivalence. These results are well known for the special case of Fermi normal coordinates, and we show that up to the second order there is no difference between the extended and the normal Fermi coordinates. In section~\ref{s:ref:ext}, we extend arbitrarily the coordinate transformations up to the third order; we derive the metric up to the second order and show that the gravitational terms cannot be canceled at this order, showing the intrinsic tidal nature of the gravitational field in a local frame. In section~\ref{sec:three} we find the link between the Fermi normal and extended coordinates, and in section~\ref{sec:four} we give some concrete examples of coordinate transformations to extended Fermi coordinates for several spacetimes.


\section{Notations and conventions} \label{s:not}
In this work the signature of the Lorentzian metric $\w{g}$ is $(+,-,-,-)$. We use natural units where the speed of light $c=1$. Greek indices run from 0 to 3 and Latin indices run from 1 to 3. The partial derivative of $A$ will be noted $A_{,\alpha}=\partial A / \partial x^\alpha$. We use the summation rule on repeated indices (one up and one down). $\eta_{\alpha \beta}$ are the components of the Minkowski metric. The convention for the Riemann tensor is:
\be \nonumber R\indices{^\mu_{\alpha \nu \beta}} = \Gamma\indices{^\mu_{\alpha \nu ,\beta}} - \Gamma\indices{^\mu _{\alpha \beta ,\nu}} - \Gamma\indices{^\mu _{\nu \sigma}} \Gamma\indices{^\sigma _{\alpha \beta}} + \Gamma\indices{^\mu _{\beta \sigma}} \Gamma\indices{^\sigma _{\alpha \nu}} \ee

The indices for tensor components in the extended Fermi frame are denoted with a hat, i.e. $\w{A} = A^{\hal} \w{e}_{\hal}$, as well as the partial derivation in the Fermi frame, i.e. $A\indices{^{\hal} _{,\hj}} = \partial A\indices{^{\hal}} / \partial X^{\hj}$.

We use the parentheses on two indices for a symmetrization  on these two indices:
\bea \nonumber A_{(\alpha \beta)} &=& \frac{1}{2} ( A_{\alpha \beta} + A_{\beta \alpha} ),\\
A_{(\alpha)} B_{(\beta)} &=& \frac{1}{2} ( A_{\alpha} B_{\beta} + A_{\beta} B_{\alpha} )  . \nonumber \eea
The indice $(\alpha \beta \nu)$ indicates even permutations of $(\alpha \beta \nu)$:
\be \nonumber (A_{\alpha \beta \nu})_{(\alpha \beta \nu)} = A_{\alpha \beta \nu} + A_{\nu \alpha \beta} + A_{\beta \nu \alpha}. \ee
We use the Levi-Civita symbol with three and four indices defined by:
\be \nonumber
\varepsilon_{\hi \hj \hk} =
\{
\begin{array}{ll}
+1 & \text{if } \ (\hi, \hj, \hk) \ \text{ is an even permutation of } (1,2,3) \\
-1 & \text{if } \ (\hi, \hj, \hk) \ \text{ is an odd permutation of } (1,2,3) \\
0 & \text{if any two labels are the same}
\end{array}
\right.
\ee
and
\be \nonumber
\varepsilon_{\hal \hbe \hat{\rho} \hat{\theta}} =
\{
\begin{array}{ll}
+1 & \text{if } \ (\hal, \hbe, \hat{\rho}, \hat{\theta}) \ \text{ is an even permutation of } (0,1,2,3) \\
-1 & \text{if } \ (\hal, \hbe, \hat{\rho}, \hat{\theta}) \ \text{ is an odd permutation of } (0,1,2,3) \\
0 & \text{if any two labels are the same}
\end{array}
\right.
\ee
%
\section{Toward extended Fermi coordinates : new approach to the second order}
\label{sec:1}
We consider the spacetime as a lorentzian manifold $(\M,\w{g})$ of dimension four. The components of the metric $g_{\mu \nu}$ are given in an \emph{initial coordinate system} $\w{x} \equiv ( x^\mu )$. $\w{x}$ is defined in an open subset $\U$. The infinitesimal interval $\dd s^2 = g_{\mu \nu} \dd x^\mu \dd x^\nu$ between two neighboring events is invariant under coordinate transformation. Let $\C$ be the observer worldline; this worldline is a timelike path. We call $s$, the proper time, that is the integral value $\int \dd s$ along $\C$ between the chosen origin $O$ and an arbitrary event $P$ along $\C$. The observer worldline is parametrized with the proper time:
\be \mathcal{C} : x^\mu = f^\mu(s) . \ee
The four-velocity is $u^\mu = \dd f^\mu / \dd s$, and the four-acceleration is $\gamma^\mu = \text{D} u^\mu / \text{D} s$, where $\text{D} / \text{D} s$ is the covariant differentiation along the worldline~$\mathcal{C}$.

We define a proper reference frame in a different way than \citeauthor{misner73}~\citep[p.327]{misner73}. It is entirely determined by these two conditions:
\begin {enumerate} 
  \item On the observer worldline, the temporal coordinate $X^{\hat{0}}$ of the proper reference frame is equal to the proper time of the observer. 
  \item At first order in the new coordinates $X^{\hal}$, we want to recover the metric of an accelerated and rotating observer in special relativity. \label{GUID2}
\end {enumerate}

The first condition comes from Fermi's idea~\citep{fermi22a}, and the second one is an extension of the usual local inertial frame condition on an accelerating and rotating observer.

The new coordinate system $\w{X} \equiv ( X^{\hal} )$ is defined in an \emph{ad-hoc} subset $\U_\C \subset \U$, so that $\C$ is included in $\U_\C$. We select $P$ an event along $\C$ so that $x^\mu_P = f^\mu(s)$. The first condition~1 infers
\be \label{phys1} X^{\hat{0}}_P = s . \ee
The origin $O$ is defined so that $x^\mu_O = f^\mu(0)$, without loss of generality. The coordinate transformation from $\w{X}$ to $\w{x}$ is a diffeomorphism $\Y : \w{X} (\U_\C) \rightarrow \w{x} (\U_\C)$.  The partial derivatives of $\Y$ at point $P$ are defined by the components of the Jacobian matrix :
\be e^\beta_{\hal} = \{ x^\beta_{,\hal} \}_P \equiv \bar{x}^\beta_{,\hal} , \ee
where $x^\beta = x^\beta (X^{\hat \alpha})$ are the components of $\Y$, and the bar stands for the value of a function at point $P$ (as $P$ is arbitrary along the worldline $\C$, then the bar stands for the value of the function all along $\C$). $e^\beta_{\hal}$ are the components of the vector $\w{e}_{\hal}$ in the natural frame associated to $\w{x}$ at event $P$: $\w{e}_{\hal} = e_{\hal}^\mu \{ \w{\partial}_{\mu} \}_P = \{ \w{\partial}_{\hal} \}_P \in \T_P (\M)$, where $\T_P$ is the tangent space at event $P$. The inverse transformations follow:
\be
\{
\begin{array}{lll}
e_\mu^{\hat{\beta}} e^\mu_{\hal} & = & \delta^{\hat{\beta}}_{\hal} \\[0.2cm]
e_\mu^{\hat{\beta}} e^\nu_{\hat{\beta}}  & = & \delta^{\nu}_{\mu} ,
\end{array}
\right.
\ee
where $\delta$ is Kronecker delta. We note that $\w{e}_{\hat{0}} = \w{u}$.

The Taylor expansion of $\Y$ at point $P$ in the hypersurface $\Sigma_s \equiv \{ Q \in \U_\C, X^{\hat{0}} (Q) = s \}$ is:
\be \label{e:ref:ccrp}
x^\mu ( X^{\hal} ) = f^\mu (s) + e^\mu_{\hj} (s) X^{\hj} + \frac{1}{2} f\indices{^\mu_{\hj \hk}} (s) X^{\hj} X^{\hk} + \Ol \left( X^3 \right) ,
\ee
where Latin letters go from 1 to 3, and $f\indices{^\mu_{\hj \hk}} (s) = \bar{x}^\mu_{,\hj \hk}$.

For the sake of simplicity, $X^{\hi} (\mathcal{C}) = 0$, ie. the worldline constitutes the spatial origin of the proper reference frame. The vector $\w{e}_{\hat{0}}$ is determined by the observer worldline, while the vectors $\w{e}_{\hj}$ are chosen so that $(\w{e}_{\hal} )$ constitutes a basis of the tangent space for each event along $\C$. $(\w{e}_{\hj} )$ is the spatial frame of the observer at event $P$.

Let $Q \in \Sigma_s$; then the Jacobian matrix of the coordinate transformation $\Y$ at event $Q$ is:
\be \label{e:ref:trans2}
\{
\begin{array}{lll}
x^\mu_{,\hat{0}} & = & u^\mu  + \dot{e}^\mu_{\hj}  X^{\hj} + \Ol \left( X^2 \right) \\[0.2cm]
x^\mu_{,\hl} & = & e^\mu_{\hl}  + f\indices{^\mu _{\hj \hl}} X^{\hj} + \Ol \left( X^2 \right) ,
\end{array}
\right.
\ee
 where $\dot{\left( \right)} \equiv \dd / \dd s$. The transformation relations of the metric tensor are $g_{\hal \hbe} = g_{\mu \nu} x^{\mu}_{,\hal} x^{\nu}_{,\hbe}$. At zeroth order it leads to
\be \label{e:ref:ordre0} \bar{g}_{\hal \hbe} = \bar{g}_{\mu \nu} e^\mu_{\hal} e^\nu_{\hat{\beta}} , \ee
where the bar stands for the value of the function at point $P$, and, at first order:
\bea
\bar{g}_{\hat{0} \hat{0} , \hj} & = &  2 \bar{g}_{\mu \nu} \dot{e}^\mu_{\hj} u^\nu + \bar{g}_{\mu \nu ,\beta} u^\mu u^\nu e^\beta_{\hj} , \label{e:ref:int1_ordre1} \\
\bar{g}_{\hat{0} \hat{m} , \hj} & = &  \left[ \bar{g}_{\mu \nu} \dot{e}^\mu_{\hj} + \left( \bar{g}_{\alpha \nu ,\mu} u^\alpha + \bar{g}_{\alpha \beta} u^\alpha f\indices{^\beta _{\mu \nu}} \right) e^\mu_{\hj} \right] e^\nu_{\hat{m} } , \label{e:ref:int2_ordre1}\\
\bar{g}_{\hat{l} \hat{m} , \hj} & = &  \left( \bar{g}_{\mu \nu, \alpha} + \bar{g}_{\sigma \nu} f\indices{^\sigma _{\alpha \mu}} + \bar{g}_{\sigma \mu} f\indices{^\sigma _{\alpha \nu}} \right) e^\mu_{\hl} e^\nu_{\hmm} e^\alpha_{\hj} , \label{e:ref:int3_ordre1}
\eea
where $f\indices{^\sigma _{\alpha \mu}}$ satisfies $f\indices{^\sigma _{\hj \hl}} = f\indices{^\sigma _{\alpha \mu}} e^\alpha_{\hj} e^\mu_{\hl}$.

\paragraph{Second order coefficients} We need a constraint in order to determine uniquely $f\indices{^\sigma _{\hj \hl}}$. The second condition~\ref{GUID2} tells us that at first order in the Fermi coordinates one should recover the special relativistic metric of an accelerated and rotating observer in local coordinates. This condition comes from the equivalence principle. According to \cite{li78}, at first order in local coordinates $(X^{\hal})$ the special relativistic metric is:
\be \nonumber \dd s^2 = \left( 1 - 2 a_{\hi} X^{\hi} \right) \dd T^2 - 2 \varepsilon_{\hj \hk \hi} \omega^{\hk} X^{\hi} \dd X^{\hj} \dd T^2 + \eta_{\hj \hmm} \dd X^{\hj} \dd X^{\hmm} ,
\ee
where $a^{\hi}$ and $\omega^{\hk}$ are respectively the observer acceleration and rotation in the local frame, and $\varepsilon_{\hi \hk \hj}$ is the Levi-Civita symbol defined in section~\ref{s:not}. This form of the metric leads to assert:
\be \label{e:ref:idee2} \bar{g}_{\hl \hmm, \hj} = 0 , \ee
\be \label{e:ass2}\bar{g}_{\hat{0} \hmm, \hj} = - \bar{g}_{\hat{0} \hj, \hmm} .  \ee

These two assertions, based on the equivalence principle, are sufficient to fix the second order coefficients $f\indices{^\sigma _{\hj \hk}}$ (see appendix~\ref{ap:one} for a demonstration):
\be \label{e:ref:coeff3} f\indices{^\sigma _{\hj \hk}} = - e^\mu_{\hj}  e^\nu_{\hk}  \bar{\Gamma}\indices{ ^\sigma _{\mu \nu}} . \ee

Now we simplify the metric coefficients in the new coordinate system. From $\nabla_\beta g_{\mu \nu} = 0$ we get:
\be \label{e:ref:def1} \bar{g}_{\mu \nu ,\beta} = \bar{\Gamma}_{\mu \nu \beta} + \bar{\Gamma}_{\nu \mu \beta} , \ee
where $\bar{\Gamma}_{\mu \nu \beta} = \bar{g}_{\alpha \mu} \bar{\Gamma}\indices{^\alpha_{\nu \beta}}$. Simplifying equations~(\ref{e:ref:int1_ordre1}) and~(\ref{e:ref:int2_ordre1}) with the relations~\eqref{e:ref:epoint} and~\eqref{e:ref:def1} we obtain:
\bea
\bar{g}_{\hat{0} \hat{0} , \hj} & = & 2 \bar{g}_{\mu \nu} \dfrac{\text{D} e^\mu_{\hj}}{\text{D} s} u^\nu ,  \label{e:ref:int2_ordre1bis} \\
\bar{g}_{\hat{0} \hat{m} , \hj} & = & \bar{g}_{\mu \nu} \dfrac{\text{D} e^\mu_{\hj}}{\text{D} s} e^\nu_{\hat{m}} . \label{e:ref:int3_ordre1bis} 
\eea
We define the antisymmetric quantity along the worldline $\C$:
\be \label{e:ref:omega}
\Omega_{\hal \hat{\beta}} = \frac{1}{2} \bar{g}_{\mu \nu} \left( e^\mu_{\hal} \frac{\text{D} e^\nu_{\hat{\beta}}}{\text{D} s} - e^\mu_{\hat{\beta}} \frac{\text{D} e^\nu_{\hal}}{\text{D} s} \right) .
\ee
Then (\ref{e:ref:int2_ordre1bis}) and (\ref{e:ref:int3_ordre1bis}) have the simple form:
\be \label{e:ref:ordre1}
\{
\begin{array}{lll}
\bar{g}_{\hat{0} \hat{0}, \hj} & = & - 2 \gamma_{\hj} \\[0.2cm]
\bar{g}_{\hat{0} \hat{m}, \hj} & = & \Omega_{\hat{m} \hj} ,
\end{array} 
\right.
\ee
where  $\gamma_{\hj} = \Omega_{\hj \hat{0}} = \bar{g}_{\mu \nu} e^\mu_{\hj} \gamma^\nu$. From the equation~(\ref{e:ref:omega}), one can deduce that the function $\Omega_{\hal \hat{\beta}}$ defines the tetrad transport along the observer trajectory:
\be \label{e:ref:trans}
\frac{\text{D} e^\mu_{\hat{\beta}}}{\text{D} s} = \Omega\indices{^\hal _\hbe} e^\mu_{\hal} ,
\ee
where $\Omega\indices{^\hal _\hbe} = \bar{g}^{\hat{\sigma} \hal} \Omega_{\hat{\sigma} \hbe}$.


If we choose the vectors $\w{e}_{\hal}$ so that they form an orthonormal basis, ie. $\bar{g}_{\hal \hbe} = \eta_{\hal \hbe}$, following equations~(\ref{e:ref:ordre0}), (\ref{e:ref:idee2}) and~(\ref{e:ref:ordre1}), we obtain a simple form of the metric in the proper reference frame, similar to the one in~\citep{misner73}:
\be \label{e:ref:metric_rp}
\begin{array}{lll}
\dd s^2 & = & \left[ 1 - 2 \gamma_{\hj} X^{\hj} + \Ol \left( X^2 \right) \right] \dd T^2 \\[0.2cm]
     &  & + \left[ 2 \Omega_{\hat{m} \hj} X^{\hj} + \Ol \left( X^2 \right) \right] \dd X^{\hat{m}} \dd T \\[0.2cm]
     &  & + \left[ \eta_{\hl \hat{m}} + \Ol \left( X^2 \right) \right] \dd X^{\hl} \dd X^{\hat{m}} .
\end{array}
\ee

\paragraph{\textbf{Fermi-Walker transport}} We can express $\Omega_{\alpha \beta}$ in terms of its Fermi-Walker part and a purely spatial rotation. Using $e^{\hat{0}}_\alpha = u_\alpha$ and $e^{\hj}_\alpha \gamma_{\hj} = \gamma_\alpha$, we deduce:
\be \label{e:ref:fw1}
\Omega_{\alpha \beta} = e_\alpha^{\hat\mu} e_\beta^{\hat\nu} \Omega_{\hat\mu \hat\nu} = \gamma_\alpha u_\beta - \gamma_\beta u_\alpha + e_\alpha^{\hi} e_\beta^{\hj} \Omega_{\hi \hj}
\ee
We define the vector $\Omega^{\hk}$ so that $\Omega_{\hi \hj} = \varepsilon_{\hi \hj \hk} \Omega^{\hk}$, with $\varepsilon_{\hi \hj \hk}$ the Levi-Civita symbol defined in section~\ref{s:not}. Then~(\ref{e:ref:fw1}) becomes:
\be \label{e:ref:fw2}
\Omega_{\alpha \beta} = \gamma_\alpha u_\beta - \gamma_\beta u_\alpha + \varepsilon_{\lambda \alpha \beta \sigma} \Omega^\sigma u^\lambda ,
\ee
where $\Omega^\sigma = e^\sigma_{\hk} \Omega^{\hk}$,  $\varepsilon_{\lambda \mu \nu \sigma} = \varepsilon_{\hat\theta \hal \hbe \hat\rho} e^{\hat\theta}_\lambda e^{\hal}_\mu e^{\hbe}_\nu e^{\hat\rho}_\sigma$ and $\varepsilon_{\hal \hbe \hat\rho \hat\theta}$ is defined in section~\ref{s:not}. We used the fact that $\w{u} = \w{e}_{\hat{0}}$ so that $\varepsilon_{\hi \hj \hk} = \varepsilon_{ \hat{\theta} \hi \hj \hk} u^{\hat{\theta}}$.

$\Omega^{\hk}$ is the rotation of the observer spatial frame $(\w{e}_{\hj})$, as it can be measured with three gyroscopes. $\gamma^{\hk}$ is the acceleration vector of the observer, as it can be measured with accelerometers. If $\Omega^{\hk}=0$ the frame is Fermi-Walker transported; if $\Omega^{\hk}=0$ and $\gamma^{\hk}=0$ the frame is parallel transported (ie. $\C$ is a geodesic).
\section{Extended Fermi coordinates : the third order} \label{s:ref:ext}
We have seen that up to the first order, the proper reference frame permits to recover the special relativity metric, as it has already been shown in \cite{misner73}. Up to the second order, some corrections are needed due to the curvature of space. These corrections depend on how one chooses to extend the spatial coordinate lines crossing the spatial origin of the Fermi reference frame. Up to now, the guideline was to recover the special relativity spacetime. \citeauthor{synge60}~\citep[p.84]{synge60}, and later \citet{manasse63}, chose geodesics to extend the spatial coordinate lines crossing the spatial origin. But as Synge himself noticed \citep[p.85]{synge60}, ``from a physical standpoint, (this choice) is somewhat artificial''. It is mathematically the simplest way to obtain a not too complicated expression of the metric. But if one extends the frame in a different way, the curvature corrections to the special relativity metric can be different. Marzlin \cite{marzlin94c} proposed a different extension from the conventional one, based on physical considerations, and proposed an experiment to expose differences between the possible coordinate systems.

We calculate now a general form of the curvature corrections, extending the Taylor's expansion of the coordinate transformation $\Y$ to the third order:
\be \label{e:ref:ccrf}
x^\mu ( X^{\hal} ) = f^\mu (s) + e^\mu_{\hj} (s) X^{\hj} + \frac{1}{2} f\indices{^\mu_{\hj \hk}} (s) X^{\hj} X^{\hk} + \frac{1}{6} f\indices{^\mu_{\hj \hk \hl}} (s) X^{\hj} X^{\hk} X^{\hl} + \Ol \left( X^4 \right) ,
\ee
where $f\indices{^\mu_{\hj \hk}}$ is given by equation~(\ref{e:ref:coeff3}), and $f\indices{^\mu_{\hj \hk \hl}} (s) = \bar{x}^\mu_{,\hat{\jmath} \hk \hl}$.

A straightforward calculation gives, at the event $Q \in \Sigma_s$:
\be \nonumber
\{
\begin{array}{lll}
x^\mu_{,\hat{0}} & = & u^\mu + \dot{e}^\mu_{\hj} X^{\hj} + \frac{1}{2} \dot{f}\indices{^\mu_{\hj \hk}} X^{\hj} X^{\hk} + \Ol \left( X^3 \right) \\[0.2cm]
x^\mu_{,\hl} & = & e^\mu_{\hl} + f\indices{^\mu_{\hj \hl}} X^{\hj} + \frac{1}{2} f\indices{^\mu_{\hj \hk \hl}} X^{\hj} X^{\hk} + \Ol \left( X^3 \right) .
\end{array}
\right.
\ee

We can underline the fact that the perturbations of gravitation on the metric components depend directly on the choice of the expansion coefficient. In order to give general relations, we keep arbitrary the functions $f\indices{^\mu_{\hj \hk \hl}}$.

We search for the metric second order terms. We can still write $g_{\hal \hat{\beta}} = g_{\mu \nu} x^\mu_{,\hal} x^\nu_{,\hat{\beta}}$, with now
\be g_{\mu \nu} \left( X^{\hal} \right) = \bar{g}_{\mu \nu} + \bar{g}_{\mu \nu ,\alpha} e^\alpha_{\hj} X^{\hj} + \frac{1}{2} \left( \bar{g}_{\mu \nu ,\alpha \beta} e^\alpha_{\hj} e^\beta_{\hk} + \bar{g}_{\mu \nu ,\alpha} f\indices{^\alpha_{\hj \hk}} \right) X^{\hj} X^{\hk} + \Ol \left( X^3 \right) . \ee

\paragraph{\textbf{The temporal part}} From transformation formulae, spatial derivatives of metric temporal components can be written along $\mathcal{C}$:
\bea \label{e:ref:inter1}
\bar{g}_{\hat{0} \hat{0}, \hj \hk} &=& 2 \bar{g}_{\mu \nu} \left( \dot{e}^\mu_{\hj} \dot{e}^\nu_{\hk} + u^\nu \dot{f}\indices{^\mu _{\hj \hk}}  \right) \nonumber \\
 &+& \bar{g}_{\mu \nu ,\beta} \left( 4 u^\mu e^\beta_{\hj} \dot{e}^\nu_{\hk}  + u^\mu u^\nu f\indices{^\beta _{\hj \hk}}  \right) \nonumber \\
 &+& \bar{g}_{\mu \nu ,\alpha \beta}  u^\mu u^\nu e^\alpha_{\hj} e^\beta_{\hk}  .
\eea
From relation~(\ref{e:ref:coeff3}) one deduces :
\be \label{e:ref:def4}
- \dot{f}\indices{^\mu _{\hj \hk}} = \dot{e}^\alpha_{\hj} e^\beta_{\hk} \bar{\Gamma}\indices{^\mu _{\alpha \beta}} + e^\alpha_{\hj} \dot{e}^\beta_{\hk} \bar{\Gamma}\indices{^\mu _{\alpha \beta}} + e^\alpha_{\hj} e^\beta_{\hk} u^\nu \bar{\Gamma}\indices{^\mu _{\alpha \beta , \nu}} .
\ee
Simplifying equation~\eqref{e:ref:inter1} with relations~\eqref{e:ref:coeff3}, \eqref{e:ref:def1} and~\eqref{e:ref:def4}, and symmetrizing the expression with $(\hj \hk)$ we obtain:
\bea \label{e:ref:inter2}
\bar{g}_{\hat{0} \hat{0}, \hj \hk} & = & 2 \bar{g}_{\mu \nu} \dot{e}^\mu_{\hj} \dot{e}^\nu_{\hk} 
+ 2 \left( \dot{e}^\alpha_{\hj} e^\beta_{\hk} + \dot{e}^\alpha_{\hk} e^\beta_{\hj} \right) \bar{\Gamma}\indices{_{\alpha \nu \beta}} u^\nu \nonumber \\
&+& e^\alpha_{\hj} e^\beta_{\hk} u^\mu u^\nu \left( \bar{g}_{\mu \nu ,\alpha \beta} - 2 \bar{\Gamma}\indices{_{\nu \alpha \beta, \mu}} - \bar{g}_{\mu \nu ,\sigma} \bar{\Gamma}\indices{^\sigma _{\alpha \beta}} \right) ,
\eea
where $\bar{\Gamma}\indices{_{\nu \alpha \beta, \mu}} = g_{\nu \sigma} \bar{\Gamma}\indices{^\sigma _{\alpha \beta, \mu}}$. Using relations~\eqref{e:ref:epoint} and \eqref{e:ref:def1} into~\eqref{e:ref:inter2} we get:
\bea \label{e:ref:inter10}
\bar{g}_{\hat{0} \hat{0}, \hj \hk } & = & 2 \bar{g}_{\mu \nu} \dfrac{\text{D} e^\mu_{\hj}}{\text{D}s} \dfrac{\text{D} e^\nu_{\hk}}{\text{D}s} \nonumber \\
 &+& e^\alpha_{\hj} e^\beta_{\hk} u^\mu u^\nu \left( \bar{g}_{\mu \nu ,\alpha \beta} - 2 \bar{\Gamma}\indices{_{\nu \alpha \beta, \mu}} - 2 \bar{\Gamma}\indices{_{\sigma \mu \alpha}} \bar{\Gamma}\indices{^\sigma _{\nu \beta}} - 2 \bar{\Gamma}\indices{_{\mu \nu \sigma}} \bar{\Gamma}\indices{^\sigma _{\alpha \beta}} \right)
\eea
From the relation $\nabla_\alpha \nabla_\beta g_{\mu \nu} = 0$ along $\C$ we get:
\be \label{e:ref:def2}
\bar{g}_{\mu \nu ,\alpha \beta} = \bar{\Gamma}_{\mu \nu \beta ,\alpha} + \bar{\Gamma}_{\nu \mu \beta ,\alpha} + \bar{g}_{\sigma \nu ,\alpha} \bar{\Gamma}\indices{^\sigma _{\beta \mu}} + \bar{g}_{\sigma \mu ,\alpha} \bar{\Gamma}\indices{^\sigma _{\beta \nu}}  .
\ee
Finally, simplifying equation~\eqref{e:ref:inter10} with~\eqref{e:ref:trans} and~\eqref{e:ref:def2} we obtain:
\be \label{e:ref:metric2_1}
\bar{g}_{\hat{0} \hat{0}, \hj \hk} = 2 \left( \Omega_{\hal \hj} \Omega\indices{^{\hal} _{\hk}} + \bar{R}_{\hat{0} \hj \hat{0} \hk} \right) ,
\ee
where $\bar{R}_{\hat{0} \hj \hat{0} \hk} = \bar{R}_{\mu \alpha \nu \beta} u^\mu u^\nu e^\alpha_{\hj} e^\beta_{\hk}$. One can remark that the temporal part does not depend on the functions $f\indices{^\alpha_{\hj \hk \hl}}$, ie. on the arbitrary part.

\paragraph{\textbf{The cross part}} The spatial derivative along $\mathcal{C}$ of the metric cross components reads:
\bea \label{e:ref:inter3}
\bar{g}_{\hat{0} \hat{m}, \hj \hk} & = & \bar{g}_{\mu \nu} \left( 2 \dot{e}^\mu_{\hj} f\indices{^\nu _{\hk \hat{m}}} + e^\nu_{\hat{m}} \dot{f}\indices{^\mu _{\hj \hk}} + u^\mu f\indices{^\nu _{\hj \hk \hat{m}}} \right) \nonumber \\
 &+& \bar{g}_{\mu \nu ,\beta} \left( 2 e^\beta_{\hj} u^\mu f\indices{^\nu _{\hk \hat{m}}} + 2 e^\beta_{\hj} \dot{e}^\mu_{\hk} e^\nu_{\hat{m}} + u^\mu e^\nu_{\hat{m}} f\indices{^\beta _{\hj \hk}} \right) \nonumber \\
 &+& \bar{g}_{\mu \nu ,\alpha \beta} e^\alpha_{\hj} e^\beta_{\hk} u^\mu e^\nu_{\hat{m}}  .
\eea
We define $f_{\mu \hj \hk \hat{m}}$ and $f_{\mu \alpha \beta \nu}$ such that $f_{\mu \hj \hk \hat{m}} = \bar{g}_{\mu \nu} f\indices{^\nu _{\hj \hk \hat{m}}} = f_{\mu \alpha \beta \nu} e^\alpha_{\hj} e^\beta_{\hk} e^\nu_{\hat{m}}$. Using the relations~(\ref{e:ref:coeff3}), (\ref{e:ref:def1}) and (\ref{e:ref:def4}), and symmetrizing equation~\eqref{e:ref:inter3} with indices $(\hj \hk)$ we obtain\footnote{We defined the symmetrization on two indices: $A_{(\alpha \beta)} = \frac{1}{2} ( A_{\alpha \beta} + A_{\beta \alpha} )$, and $A_{(\alpha)} B_{(\beta)} = \frac{1}{2} ( A_{\alpha} B_{\beta} + A_{\beta} B_{\alpha} )$}:
\bea \label{e:ref:inter11}
\bar{g}_{\hat{0} \hat{m}, \hj \hk} & = & e^\alpha_{\hj} e^\beta_{\hk} e^\nu_{\hat{m}} u^\mu \left( f_{\mu \alpha \beta \nu} + \bar{g}_{\mu \nu ,\alpha \beta} \right. \nonumber \\
 &-&  \left.  \bar{\Gamma}_{\nu \alpha \beta ,\mu} - \bar{g}_{\mu \nu ,\sigma} \bar{\Gamma}\indices{^\sigma _{\alpha \beta}} - 2 \bar{g}_{\mu \sigma , ( \alpha ) } \bar{\Gamma}\indices{^\sigma _{ ( \beta ) \nu}} \right) ,
\eea
From the relation $\nabla_\alpha \nabla_\beta g_{\mu \nu} = 0$ along $\C$ we get:
\be \label{e:ref:def5}
\bar{g}_{\mu \nu ,\alpha \beta} = \bar{\Gamma}_{\mu \nu ( \alpha ,\beta )} + \bar{\Gamma}_{\nu \mu ( \alpha ,\beta )} + \bar{g}_{\sigma \nu , (\alpha)} \bar{\Gamma}\indices{^\sigma _{(\beta ) \mu}} + \bar{g}_{\sigma \mu , ( \alpha)} \bar{\Gamma}\indices{^\sigma _{(\beta ) \nu}}  .
\ee
Simplifying~\eqref{e:ref:inter11} with~(\ref{e:ref:def1}) and \eqref{e:ref:def5} we obtain:
\bea \label{e:ref:inter12}
\bar{g}_{\hat{0} \hat{m}, \hj \hk} & = & e^\alpha_{\hj} e^\beta_{\hk} e^\nu_{\hat{m}} u^\mu \left( f_{\mu \alpha \beta \nu} - \bar{R}\indices{_{\nu ( \alpha \beta ) \mu}} \right. \nonumber \\
 &+& \left. \bar{\Gamma}\indices{_{\mu \nu ( \alpha , \beta )}} - \bar{\Gamma}\indices{_{\mu \sigma ( \alpha ) }} \bar{\Gamma}\indices{^\sigma _{ ( \beta ) \nu}} - \bar{\Gamma}\indices{_{\mu \sigma \nu}} \bar{\Gamma}\indices{^\sigma _{\alpha \beta}} \right) .
\eea
We define\footnote{The indice $(\alpha \beta \nu)$ indicates even permutations of $(\alpha \beta \nu)$~: $(A_{\alpha \beta \nu})_{(\alpha \beta \nu)} = A_{\alpha \beta \nu} + A_{\nu \alpha \beta} + A_{\beta \nu \alpha}$.}:
\be \label{e:ref:phi}
\phi_{\mu \alpha \beta \nu} = f_{\mu \alpha \beta \nu} + \left( \frac{1}{3} \bar{\Gamma}_{\mu \nu \alpha ,\beta} - \frac{2}{3} \bar{\Gamma}_{\mu \sigma \nu} \bar{\Gamma}\indices{^\sigma _{\alpha \beta}} \right)_{(\alpha \beta \nu)} ,
\ee
This quantity is symmetrical with respect to~$(\alpha \beta \nu)$. Finally, using the definition~\eqref{e:ref:phi} into equation~\eqref{e:ref:inter12} we obtain the simple form:
\be \label{e:ref:metric2_2}
\bar{g}_{\hat{0} \hmm, \hj \hk} = - \frac{4}{3} \bar{R}_{\hat{0} ( \hj \hk ) \hmm} + \phi_{\hat{0} \hj \hk \hmm} ,
\ee
with $\bar{R}_{\hat{0} \hj \hk \hmm} = \bar{R}_{\mu \alpha \beta \nu} u^\mu e^\alpha_{\hj} e^\beta_{\hk} e^\nu_{\hmm}$ and $\phi_{\hat{0} \hj \hk \hmm} = \phi_{\mu \alpha \beta \nu} u^\mu e^\alpha_{\hj} e^\beta_{\hk} e^\nu_{\hmm}$.

\paragraph{\textbf{The spatial part}} Spatial derivatives of the metric spatial components can be obtained from the transformation relations along $\mathcal{C}$:
%
\bea \label{e:ref:inter5}
\bar{g}_{\hl \hmm, \hj \hk} & = & \bar{g}_{\mu \nu} \left( 2 f\indices{^\mu _{\hj \hl}} f\indices{^\nu _{\hk \hmm}} + e^\nu_{\hmm} f\indices{^\mu _{\hj \hk \hl}}  + e^\mu_{\hl} f\indices{^\nu _{\hj \hk \hmm}}  \right) \nonumber \\
 &+& \bar{g}_{\mu \nu ,\beta}  \left( 2 e^\beta_{\hj} e^\mu_{\hl} f\indices{^\nu _{\hk \hmm}}  + 2 e^\beta_{\hj} e^\nu_{\hmm} f\indices{^\mu _{\hk \hl}}  + e^\mu_{\hl} e^\nu_{\hmm} f\indices{^\beta _{\hj \hk}}  \right) \nonumber \\
 &+& \bar{g}_{\mu \nu ,\alpha \beta} e^\alpha_{\hj} e^\beta_{\hk} e^\mu_{\hl} e^\nu_{\hmm} ,
\eea
Symmetrizing this expression with indices $(\hj \hk)$ and $(\hl \hmm)$ and using relation~\eqref{e:ref:coeff3} we find:
\bea \label{e:ref:inter15}
\bar{g}_{\hl \hmm, \hj \hk} & = & e^\alpha_{\hj} e^\beta_{\hk} e^\mu_{\hl} e^\nu_{\hmm} \left( 2 f\indices{_{( \mu \nu ) \alpha \beta}} + \bar{g}_{\mu \nu ,\alpha \beta} + 2 \bar{\Gamma}\indices{_{\sigma \alpha (\mu)}} \bar{\Gamma}\indices{^\sigma _{(\nu ) \beta}} \right. \nonumber\\
 &-& \left. 2 g\indices{_{\mu \sigma, (\alpha)}} \bar{\Gamma}\indices{^\sigma _{(\beta ) \nu}} - 2 g\indices{_{\nu \sigma, (\alpha)}} \bar{\Gamma}\indices{^\sigma _{(\beta ) \mu}} - g\indices{_{\mu \nu, \sigma}} \bar{\Gamma}\indices{^\sigma _{\alpha \beta}} \right)
\eea
Then, symplifying~\eqref{e:ref:inter15} with~\eqref{e:ref:def1}, \eqref{e:ref:def5} and~\eqref{e:ref:phi} we obtain:
\be \label{e:ref:metric2_3}
\bar{g}_{\hl \hmm, \hj \hk} = - \frac{1}{3} \left( \bar{R}_{\hl ( \hj \hk ) \hmm} + \bar{R}_{\hk ( \hl \hmm ) \hj} \right) + 2\phi_{ ( \hl \hmm ) \hj \hk} , 
\ee
where $\bar{R}_{\hl \hj \hmm \hk} = \bar{R}_{\mu \alpha \nu \beta} e^\mu_{\hl} e^\alpha_{\hj} e^\nu_{\hmm} e^\beta_{\hk}$ and $\phi_{\hl \hmm \hj \hk} = \phi_{\mu \alpha \beta \theta} e^\mu_{\hl} e^\alpha_{\hmm} e^\beta_{\hat{j}} e^\theta_{\hk}$.

\paragraph{\textbf{The metric}} The metric of the extended Fermi reference frame is deduced from equations~(\ref{e:ref:metric_rp}), (\ref{e:ref:metric2_1}), (\ref{e:ref:metric2_2}) and~(\ref{e:ref:metric2_3}):
\be \label{e:ref:metric_cf}
\begin{array}{lll}
\dd s^2 & = & \left[ 1 - 2 \gamma_{\hj} X^{\hj} + \left( \Omega_{\hal \hj} \Omega\indices{^{\hal} _{\hk}} + \bar{R}_{\hat{0} \hj \hat{0} \hk} \right) X^{\hj} X^{\hk} + \Ol \left( X^3 \right) \right] \dd T^2 \\[0.2cm]
     & & + \left[ 2 \Omega_{\hmm \hj} X^{\hj} + \left( \frac{4}{3} \bar{R}_{\hat{0} \hj \hmm \hk} + \phi_{\hat{0} \hj \hk \hmm} \right) X^{\hj} X^{\hk} + \Ol \left( X^3 \right) \right] \dd X^{\hmm} \dd T \\[0.2cm]
     & & + \left[ \eta_{\hl \hmm} + \left( \frac{1}{3} \bar{R}_{\hl \hj \hmm \hk} + \phi_{\hl \hmm \hj \hk} \right) X^{\hj} X^{\hk} + \Ol \left( X^3 \right) \right] \dd X^{\hl} \dd X^{\hmm} ,
\end{array}
\ee
where  $\phi_{\hat{\sigma} \hmm \hj \hk} = \phi_{\mu \alpha \beta \theta} e^\mu_{\hat{\sigma}} e^\alpha_{\hmm} e^\beta_{\hat{j}} e^\theta_{\hk}$, and $\phi_{\mu \alpha \beta \theta}$ is defined by (\ref{e:ref:phi}).

Can we cancel in~\eqref{e:ref:metric_cf} the second order terms for the crossed coefficients of the metric: $\bar{g}_{\hat{0} \hmm, \hj \hk} X^{\hj} X^{\hk}$ and/or for the spatial coefficients: $\bar{g}_{\hl \hmm, \hj \hk} X^{\hj} X^{\hk}$, with a good choice of the ``$\phi$-terms''? This is impossible. Let's prove it for the spatial coefficients:
\bea \bar{g}_{\hl \hmm, \hj \hk} X^{\hj} X^{\hk} = 0 
	& \Leftrightarrow & \left( \frac{1}{6} \left( \bar{R}_{\hl \hj \hmm \hk} + \bar{R}_{\hl \hk \hmm \hj} \right) + \phi_{\hl \hmm \hj \hk} \right) X^{\hj} X^{\hk} = 0 \nonumber \\
	& \Leftrightarrow & \phi_{\hl \hmm \hj \hk} = - \frac{1}{6} \left( \bar{R}_{\hl \hj \hmm \hk} + \bar{R}_{\hl \hk \hmm \hj} \right) \nonumber \\
	& \Rightarrow & \phi_{\hl \hmm \hj \hk} = 0 \nonumber ,
\eea
where for the final line we symmetrized the expression with the indices $(\hj \hmm \hk)$. The demonstration is similar for the crossed coefficients of the metric. It is therefore impossible to cancel the curvature terms at the second order with a good choice of the ``$\phi$-terms''. This shows the intrinsic tidal nature of the gravitational field in a local frame.

It is interesting to remark that the temporal part of the metric does not depend on $f\indices{^\mu _{\hj \hk \hl}}$. It means that the physical assumptions \eqref{phys1}, \eqref{e:ref:idee2} and~\eqref{e:ass2} are sufficient to set the temporal form of the extended Fermi metric up to the second order. We emphasize that the second order terms do not depend on the choice of the initial coordinate system. Indeed, from the point of view of a change of the initial coordinate system, $\phi_{\hat{\sigma} \hat{\jmath} \hat{k} \hat{m}}$ is a scalar. Then the gravitational corrections depend only on the choice of the extended Fermi coordinates.
\section{Extended Fermi coordinates and the Fermi normal coordinates} \label{sec:three}
The general coordinate transformations from the initial coordinates to the extended Fermi coordinates are
\begin{equation}
x^\mu = f^\mu + e^\mu_{\hat{\jmath}} X^{\hat{\jmath}} - \dfrac{1}{2} \bar{\Gamma}\indices{^\sigma _{\mu \nu}} e^\mu_{\hat{\jmath}} e^\nu_{\hat{k}} X^{\hat{\jmath}} X^{\hat{k}} + \frac{1}{6} f\indices{^\mu_{\hat{\jmath} \hat{k} \hat{l}}} X^{\hat{\jmath}} X^{\hat{k}} X^{\hat{l}} + O \left( X^4 \right)
\end{equation}
\begin{figure}[t]
\centerline{\includegraphics[width=0.8\linewidth]{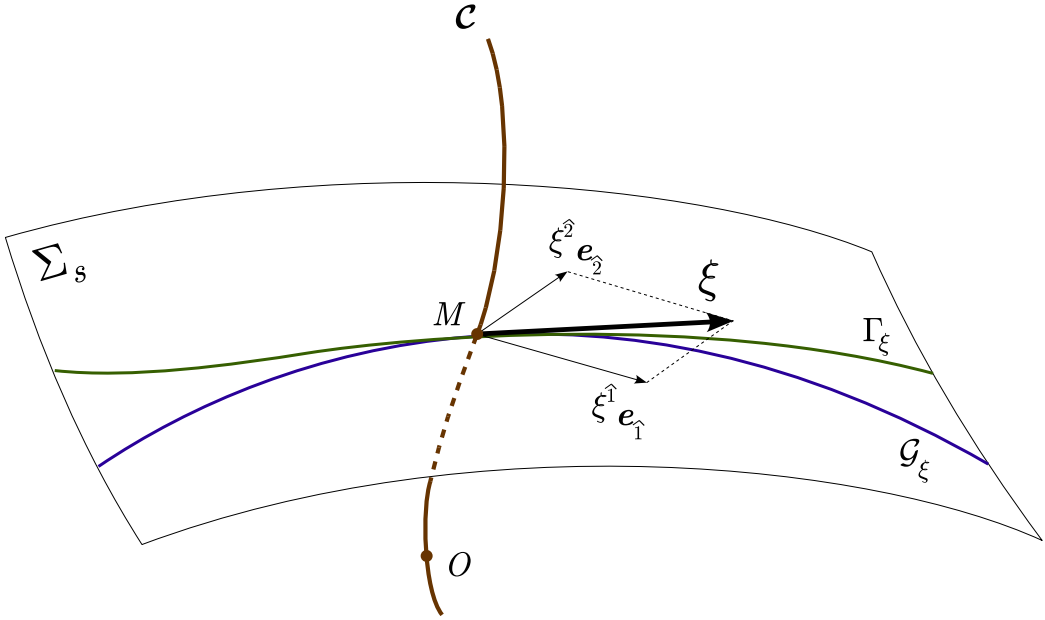}}
\caption[]{\label{e:ref:trans-FIG} \footnotesize 2D illustration of a spatial coordinate line of the proper reference frame crossing the observer worldline $\C$, in the hypersurface $\Sigma_s$. The curve $\mathcal{G}_\xi$, a spatial geodesic, is obtained with a parallel transport of vectors $\w{a}_{\hal}$. $\Gamma_\xi$ is obtained with an arbitrary extension of the transport of vectors $\w{a}_{\hal}$. $\w{\xi}$ is the common tangent vector to the curves $\mathcal{G}_\xi$ et $\Gamma_\xi$ at point $P$.}
\end{figure}
From this equation one can easily deduce the equations of the Fermi reference frame coordinate lines in the initial coordinates. The choice of the functions $f\indices{^\mu _{\hat{\jmath} \hat{k} \hat{l}}}$ determines how one extends the spatial coordinate lines which cross point $P$. 

We will now find the functions $f\indices{^\mu _{\hj \hk \hl}}$ for the Fermi normal coordinates. In these coordinates, the spatial coordinate lines crossing the observer worldline $\mathcal{C}$ are geodesics~(see figure~\ref{e:ref:trans-FIG}). This choice is not based on physical assumptions. It is equivalent to say that the vectors $\w{a}_{\hk} \equiv (x^{\mu}_{,\hk})$ are parallel transported along these spatial coordinate lines. In $\Sigma_s$, the equations of the spatial coordinate lines $\Gamma_\xi$ crossing the event $P$ are $X^{\hat{\jmath}} = \xi^{\hat{\jmath}} u$, where $u$ is a parameter and $\w{\xi}$ the spatial vector tangent to $\Gamma_\xi$ at event P. The equations of parallel transport of the vectors $\w{a}_{\hl} \equiv (x^{\mu}_{,\hl})$ along $\Gamma_\xi$ are
\begin{equation}
\label{GE}
\frac{\text{d} a^\mu_{\hat{l}}}{\text{d} u} + \Gamma\indices{^\mu _{\alpha \beta}} a^\alpha_{\hat{l}} a^\beta_{\hat{\jmath}} \frac{\text{d} X^{\hat{\jmath}}}{\text{d} u} = 0 .
\end{equation}

At zeroth order in $u$, this equation is equivalent to
\begin{equation}
\nonumber
f\indices{^\mu _{\hat{\jmath} \hat{l}}} = - \bar{\Gamma}\indices{^\mu _{\alpha \beta}} e^\alpha_{\hat{l}} e^\beta_{\hat{\jmath}} + O(u) .
\end{equation}

This is true for all extended Fermi coordinates, as shown in section~2~(eq.(\ref{e:ref:coeff3})). It means that the physical assumptions~(\ref{e:ref:idee2}) and~\eqref{e:ass2} are equivalent to the fact that there is a three-point contact between $\mathcal{G}_\xi$ and $\Gamma_\xi$ at point $P$~(see figure~\ref{e:ref:trans-FIG}).

To constrain the coefficients $f\indices{^\mu _{\hat{\jmath} \hat{k} \hat{l}}}$ we need to apply the geodesic deviation equation to $\Gamma_\xi$. For this, we define a new line in $\Sigma_s$, close to $\Gamma_\xi$, with tangent vector $(\w{\xi + \delta \xi})$ at event $P$: $\Gamma_{\xi + \delta \xi} : X^{\hj} = \left( \xi^{\hj} + \delta \xi^{\hj} \right) u$.

In the sequel, the sign $\overset{0}{\left( \right)}$ stands for the value of the functions along $\Gamma_\xi$. Along $\Gamma_{\xi + \delta \xi}$, we can write
\begin{equation}
\label{GE1}
a^\mu_{\hat{l}} = \overset{0}{a^\mu_{\hat{l}}} + \left( f\indices{^\mu _{\hat{\jmath} \hat{l}}} u + f\indices{^\mu _{\hat{\jmath} \hat{k} \hat{l}}} \xi^{\hat{k}} u^2 + O \left( u^3 \right) \right) \delta \xi^{\hat{\jmath}} + O \left( \delta \xi^2 \right) ,
\end{equation}
where $\overset{0}{a^\mu_{\hat{l}}} = e^\mu_{\hat{l}} + f\indices{^\mu _{\hat{\jmath} \hat{l}}} \xi^{\hat{\jmath}} u + O \left( u^2 \right)$. Taking the derivative with $u$ we obtain
\begin{equation}
\label{GE2}
\frac{\text{d} a^\mu_{\hat{l}}}{\text{d}u} = \frac{\text{d} \overset{0}{a^\mu_{\hat{l}}}}{\text{d}u} + \left( f\indices{^\mu _{\hat{\jmath} \hat{l}}} + 2 f\indices{^\mu _{\hat{\jmath} \hat{k} \hat{l}}} \xi^{\hat{k}} u \right) \delta \xi^{\hat{\jmath}} + O \left( \delta \xi^2 \right) .
\end{equation}

Moreover,
\begin{equation}
\label{GE3}
\Gamma\indices{^\mu _{\alpha \beta}} = \overset{0}{\Gamma\indices{^\mu _{\alpha \beta}}} + \left( \Gamma\indices{^\mu _{\alpha \beta ,\theta}} e^\theta_{\hat{\jmath}} u + O \left( u^2 \right) \right) \delta \xi^{\hat{\jmath}} + O \left( \delta \xi^2 \right),
\end{equation}
where $\overset{0}{\Gamma\indices{^\mu _{\alpha \beta}}} = \bar{\Gamma}\indices{^\mu _{\alpha \beta}} + \bar{\Gamma}\indices{^\mu _{\alpha \beta ,\theta}} e^\theta_{\hat{\jmath}} \xi^{\hat{\jmath}} u + O \left( u^2 \right)$.

Finally, introducing equations~(\ref{GE1}),~(\ref{GE2}) and~(\ref{GE3}) in the equation of the parallel transport~(\ref{GE}), and subtracting~(\ref{GE}) up to the zeroth order in $\delta \xi$, we obtain
\begin{equation}
\label{COEFF2}
f\indices{^\mu _{\hat{\jmath} \hat{k} \hat{l}}} = \frac{1}{2} \left( 2 \bar{\Gamma}\indices{^\mu _{\alpha \sigma}} \bar{\Gamma}\indices{^\sigma _{\beta \theta}} + \bar{\Gamma}\indices{^\mu _{\beta \sigma}} \bar{\Gamma}\indices{^\sigma _{\theta \alpha}} + \bar{\Gamma}\indices{^\mu _{\theta \sigma}} \bar{\Gamma}\indices{^\sigma _{\beta \alpha}} - 2 \bar{\Gamma}\indices{^\mu _{\alpha \beta ,\theta}} \right) e^\theta_{\hat{\jmath}} e^\beta_{\hat{k}} e^\alpha_{\hat{l}} .
\end{equation}

Symmetrizing this expression with $(\hj \hk \hl)$, we obtain
\begin{equation}
\left[ f\indices{^\mu _{\hat{\jmath} \hat{k} \hat{l}}} \right]_{\text{FNC}} = \left( \frac{2}{3} \bar{\Gamma}\indices{^\mu _{\theta \sigma}} \bar{\Gamma}\indices{^\sigma _{\alpha \beta}} - \frac{1}{3} \bar{\Gamma}\indices{^\mu _{\alpha \beta ,\theta}} \right)_{(\alpha \beta \theta)} e^\theta_{\hat{\jmath}} e^\beta_{\hat{k}} e^\alpha_{\hat{l}} ,
\end{equation}

where FNC stands for Fermi normal coordinates. Then, we deduce from the expression~(\ref{e:ref:phi}) that $\left[ \phi_{\mu \alpha \beta \theta} \right]_{\text{FNC}} = 0$. The metric in the FNC is the metric~(\ref{e:ref:metric_cf}) with $\phi_{\hal \hmm \hj \hk}=0$. This result is in agreement with the one of Ni \& Zimmermann \cite{ni78}. They obtained the metric with a very different method, but it does not permit to calculate the coordinate transformation from the initial coordinates to the FNC. This coordinate transformation is:
\be \label{CT}
\begin{array}{lll}
x^\mu & = & f^\mu + e^\mu_{\hj} X^{\hj} - \dfrac{1}{2} \bar{\Gamma}\indices{^\mu _{\alpha \beta}} e^\alpha_{\hj} e^\beta_{\hk} X^{\hj} X^{\hk}\\[0.2cm]
 & + & \left( \frac{1}{9} \bar{\Gamma}\indices{^\mu _{\theta \sigma}} \bar{\Gamma}\indices{^\sigma _{\alpha \beta}} - \frac{1}{18} \bar{\Gamma}\indices{^\mu _{\alpha \beta ,\theta}} \right)_{(\alpha \beta \theta)} e^\theta_{\hj} e^\beta_{\hk} e^\alpha_{\hl} X^{\hj} X^{\hk} X^{\hl} + \Ol \left( X^4 \right)
\end{array}
\ee

This result is in agreement with the one of \citeauthor{klein08}~\citep[eq.27]{klein08}. In a weak gravitational field, the coefficients $\bar{\Gamma}\indices{^\mu _{\theta \sigma}} \bar{ \Gamma } \indices{ ^\sigma _{\alpha \beta}}$ are of the second order and can be neglected. Then the formula~(\ref{CT}) is in agreement with the one found by \citeauthor{ashby86}~\citep[eq.9]{ashby86}.
\section{Examples of extended Fermi coordinates} \label{sec:four}
This section gives examples of extended coordinates. In the first example we give an extended Fermi coordinate system in order to simplify the components of the metric tensor in a Schwarzschild metric. In the second example, we introduce three extended Fermi coordinates in a de Sitter metric, in order to discuss their physical meaning.

\paragraph{\textbf{The Schwarzschild metric}}
The metric in the usual Schwarzschild coordinates $(ct,r,\theta,\phi)$ is
\be \dd s^2 = \left( 1-\dfrac{2m}{r} \right) \dd t^2 - \dfrac{1}{1-\frac{2m}{r}} \dd r^2 - r^2 \left( \dd \theta^2 + \sin^2 \theta \dd \phi^2 \right) \ee

We consider the static observer, with a worldline $\C$ located at $r=R$, $\theta = \pi/2$ and $\phi = 0$, with $R>2m$. Note that $\C$ is not a geodesic. The simplest tetrad along $\C$ is:
\be 
\w{e}_{\hat{0}} = \Delta^{-1} \w{\partial}_t \ ; \ 
\w{e}_{\hat{1}} = \Delta \w{\partial}_r \ ; \ 
\w{e}_{\hat{2}} = R^{-1} \w{\partial}_\theta \ ; \ 
\w{e}_{\hat{3}} = R^{-1} \w{\partial}_\phi
\ee
where $\Delta = \sqrt{1-\frac{2m}{R}}$.

In the Fermi normal coordinates $(X^{\hal})$, the metric can be derived from the \citeauthor{ni78} formula~\citep{ni78}. The result is, up to the second order:
\be \begin{array}{lll}  
\dd s^2 &=& \eta_{\hal \hbe} \dd X^{\hal} \dd X^{\hbe} + \left[ \dfrac{2m}{\Delta R^2} X^{\hat{1}} + \left( \dfrac{m}{\Delta R^2} X^{\hat{1}} \right)^2 + \dfrac{m}{R^3} \left( - 2 (X^{\hat{1}})^2 + (X^{\hat{2}})^2 + (X^{\hat{3}})^2 \right) \right] ( \dd X^{\hat{0}} )^2 \\[0.3cm] 
&& - \dfrac{m}{3R^3} \left[ \left( (X^{\hat{2}})^2+(X^{\hat{3}})^2 \right) (\dd X^{\hat{1}})^2 + \left( (X^{\hat{1}})^2-2(X^{\hat{3}})^2 \right) (\dd
X^{\hat{2}})^2+ \left( (X^{\hat{1}})^2-2(X^{\hat{2}})^2 \right) (\dd X^{\hat{3}})^2 \right. \\[0.3cm]
&& \left. - 2 \left( X^{\hat{1}} X^{\hat{2}} \dd X^{\hat{1}} \dd X^{\hat{2}} + X^{\hat{1}} X^{\hat{3}} \dd X^{\hat{1}} \dd X^{\hat{3}}- 2 X^{\hat{2}} X^{\hat{3}} \dd X^{\hat{2}} \dd X^{\hat{3}} \right) \right] 
\end{array} \ee

We can take advantage of the extended Fermi coordinate new degree of freedom to write the metric in a simplified form. We have already seen that it is not possible to cancel all the second order terms. But we can choose the coefficients $\phi_{\hl \hmm \hj \hk}$, in the expression~(\ref{e:ref:metric_cf}) of the metric, in order to diagonalize the metric tensor. Then, with these coefficients, we can deduce the coefficients $f\indices{^\mu _{\hj \hk \hl}}$ and calculate the coordinate transformations. For example, with the coordinate transformations
\be \{ \begin{array}{lll} ct &=& \Delta^{-1} X^{0} \equiv \Delta^{-1} X^{\hat{0}} \\[0.5cm]
r &=& R + \Delta X^1 + \dfrac{m}{2R^2}
(X^1)^2 + \dfrac{\Delta^2}{2R} \left( (X^2)^2+(X^3)^2 \right) \\[0.2cm] &&+
\dfrac{\Delta}{3R^3} \left[ -\dfrac{m}{R} (X^1)^2 + \left(
\dfrac{11 m}{2R}-\dfrac{3}{2} \right) \left( (X^2)^2+(X^3)^2
\right) \right] X^1 \\[0.5cm]
\theta &=& \dfrac{\pi}{2} + \dfrac{X^2}{R} - \dfrac{\Delta}{R^2}
X^1 X^2 + \dfrac{1}{R^3} \left[ \left( 1-\dfrac{7m}{3R} \right)
(X^1)^2 - \dfrac{\Delta^2}{3} (X^2)^2 -  \dfrac{1}{2} \left( 1-
\dfrac{m}{R} \right) (X^3)^2
\right] X^2 \\[0.5cm]
\phi &=& \dfrac{X^3}{R} - \dfrac{\Delta}{R^2} X^1 X^3 +
\dfrac{1}{R^3} \left[ \left( 1-\dfrac{7m}{3R} \right) (X^1)^2 +
\dfrac{m}{2R} (X^2)^2 -  \dfrac{\Delta^2}{3} (X^3)^2 \right] X^3
\end{array} 
\right. 
\ee
to the extended Fermi coordinates $(X^{0},X^1,X^2,X^3)$, the metric is:
\be \begin{array}{lll} \dd s^2 &=&  \eta_{\alpha \beta} \dd X^{\alpha} \dd X^{\beta} + \left[ \dfrac{2m}{\Delta R^2} X^{1} + \left( \dfrac{m}{\Delta R^2} X^{1} \right)^2 + \dfrac{m}{R^3} \left( - 2 (X^{1})^2 + (X^{2})^2 + (X^{3})^2 \right) \right] ( \dd X^{0} )^2 \\[0.3cm] 
&& - \dfrac{m}{3R^3} \left[ 2 \left(
(X^2)^2+(X^3)^2 \right) (\dd X^1)^2 + \left( (X^1)^2-3(X^3)^2
\right) (\dd X^2)^2+ \left( (X^1)^2-3(X^2)^2 \right) (\dd X^3)^2
\right]
\end{array} \ee

This result shows that the Fermi normal frame is not the simplest extended Fermi frame to write the components of the metric. The Fermi normal coordinates and the extended Fermi coordinates introduced here are equivalent up to terms of the second order in the coordinates.

\paragraph{\textbf{de Sitter metric}}
We take as initial coordinates $(x^\alpha)$ the ones of the de Sitter metric:
\be \dd s^2 = -\dd t^2 + \A^2 (t) \delta_{ij} \dd x^i
\dd x^j \ , \ \A^2(t) = e^{H t} \ (H>0) \ee

We examine the simplest trajectory for the observer:
\be x^0=s \ ; \ x^i=0 .
\ee
We introduce the normal Fermi coordinates system, $(X^{\hat{\alpha}})$, and two different extended Fermi coordinates systems: $(X^{\stackrel{\ast}{\alpha}})$ and $(X^{\stackrel{\times}{\alpha}})$. We note that $X^{\hat{0}} = X^{\stackrel{\ast}{0}} = X^{\stackrel{\times}{0}} = s$. The three Fermi coordinates systems are defined by the coordinate transformations up to the third order :
\be x^0 = s - \dfrac{H \hat{R}^2}{2} + \Ol(H^3 \hat{R}^4) = s -
\dfrac{H \stackrel{\ast}{R}^2}{2} + \Ol(H^3 \stackrel{\ast}{R}^4) = s - \dfrac{H
\stackrel{\times}{R}^2}{2} + \Ol(H^3 \stackrel{\times}{R}^4) \ee
where $R^2 \equiv \delta_{ij} X^i X^j$, and
\be \begin{array}{lll} x^i &=& e^{-Hs} \left(1+\dfrac{H^2
\hat{R}^2}{3} + \Ol(H^3 \hat{R}^3) \right) X^{\hi}\\[0.3cm]
x^i &=& e^{-Hs} \left(1+\dfrac{H^2
\stackrel{\ast}{R}^2}{4} + \Ol(H^3 \stackrel{\ast}{R}^3) \right) X^{\stackrel{\ast}{\imath}}\\[0.3cm]
x^i &=& e^{-Hs} \left(1+\dfrac{H^2 \stackrel{\times}{R}^2}{2} + \Ol(H^3
\stackrel{\times}{R}^3) \right) X^{\stackrel{\times}{\imath}} \end{array} \ee

The three radial coordinates are equivalent up to the second order, but differ at orders superior or equal to three:
\be \hat{R} = \left( 1-\dfrac{H^2 \stackrel{\ast}{R}^2}{12} + \Ol(H^3
\stackrel{\ast}{R}^3) \right) \stackrel{\ast}{R} = \left( 1-\dfrac{H^2 \stackrel{\times}{R}^2}{6} +
\Ol(H^3 \stackrel{\times}{R}^3) \right) \stackrel{\times}{R} \ee

The time and crossed components of the metric are the same, up to the second order, in the three Fermi frames:
\be \begin{array}{lllllll} g_{\hat{0} \hat{0}} &=& g_{\stackrel{\ast}{0} \stackrel{\ast}{0}}
&=& g_{\stackrel{\times}{0} \stackrel{\times}{0}} &=& -1 + H^2 R^2 + \Ol(H^3 R^3) \\[0.3cm]
g_{\hat{0} \hat{m}} &=& g_{\stackrel{\ast}{0} \stackrel{\ast}{m}} &=& g_{\stackrel{\times}{0} \stackrel{\times}{m}}
&=& 0 + \Ol(H^3 R^3)
\end{array} \ee
where $R$ is either $\hat{R}$, $\stackrel{\ast}{R}$ or $\stackrel{\times}{R}$. It is not necessary to specify which $R$ is chosen because we have shown that they are equivalent up to the second order.

On the other hand, the spatial components of the metric differ at orders superior or equal to two:
\be \begin{array}{lll} g_{\hat{l} \hat{m}} &=& \delta_{\hl \hmm}
\left( 1 - \dfrac{H^2 \hat{R}^2}{3} \right) + \dfrac{H^2}{3}
X^{\hl} X^{\hmm} + \Ol(H^3 \hat{R}^3) \\[0.5cm]
g_{\stackrel{\ast}{l} \stackrel{\ast}{m}} &=& \delta_{\stackrel{\ast}{l} \stackrel{\ast}{m}} \left( 1 -
\dfrac{H^2 \stackrel{\ast}{R}^2}{2} \right) + \Ol(H^3 \stackrel{\ast}{R}^3) \\[0.5cm]
g_{\stackrel{\times}{l} \stackrel{\times}{m}} &=& \delta_{\stackrel{\times}{l} \stackrel{\times}{m}} + H^2
X^{\stackrel{\times}{l}} X^{\stackrel{\times}{m}} + \Ol(H^3 \stackrel{\times}{R}^3) \end{array} \ee

One can note that the metric is diagonal in the extended Fermi coordinate system~$(X^{\stackrel{\ast}{\alpha}})$.
\section{Conclusion}

In this article, we have developed the formalism of extended Fermi coordinates. We demonstrated that this enlarged definition of Fermi coordinates follows the original Fermi's idea, but constitutes a generalized approach to the third order description of systems. More precisely, the equivalence principle determines the coordinate transformations up to the second order. At the third order, the relevant parameters of this description are the arbitrary functions $f^\mu _{\hat{\jmath} \hat{k} \hat{l}}$, and we calculated the expression of this function in the case of Fermi normal coordinates. We calculated the metric in the extended Fermi coordinates and have shown some properties of the second order gravitational terms: the temporal part is fixed by the physical assumptions \eqref{phys1}, \eqref{e:ref:idee2} and~\eqref{e:ass2}; the crossed and the spatial parts depend on which extended Fermi coordinates one chooses; they cannot be canceled by a good choice of the extended Fermi coordinates, which shows the intrinsic tidal nature of the gravitational field in a local frame. These assertions raise the question of the physical meaning of the Fermi coordinates, already asked by \citet{marzlin94c}.

On the one hand, if an experiment is operationally defined in a non-covariant way using Fermi coordinates, then the definition of the experiment will be different if one uses different extended Fermi coordinates, and the outcome of these experiment will be \emph{a priori} different; on the other, analyzing the outcome of an experiment in terms of Fermi coordinates will lead \emph{a priori} to different interpretation when using different extended Fermi coordinates, as shown in the examples. Only an experiment can tell which one is better suited for experiments. But, as \citet{marzlin94c} underlined it, the required precision is far beyond the scope of any experiment on Earth or even in the Solar System. Observations of strong gravitational potential effects, such as black hole vicinity tidal effects, should bring some clue to this problem, but once again, the information is presently not reachable.


\appendix
\section{The second order coefficients}
\label{ap:one}
Using the assertion~\eqref{e:ref:idee2} and equation~\eqref{e:ref:int3_ordre1} we obtain:
\be \bar{g}_{\mu \nu, \alpha} e^\mu_{\hl} e^\nu_{\hmm} e^\alpha_{\hj} + \bar{g}_{\sigma \nu} f\indices{^\sigma _{\hj \hl}} e^\nu_{\hmm} + \bar{g}_{\sigma \mu} f\indices{^\sigma _{\hj \hmm}} e^\mu_{\hl} = 0 \label{e:ass3} \ee
%
The covariant differentiation of $\w{e}_{\hk}$ gives:
\be \label{e:ref:epoint} \dot{e}^\mu_{\hk} = \frac{\text{D} e^\mu_{\hk}}{\text{D} s} - \bar{\Gamma}\indices{^\mu _{\alpha \beta}} u^\beta e^\alpha_{\hk} , \ee
where $\bar{\Gamma}\indices{^\mu _{\alpha \beta}}$ are the connection coefficients along the worldline~$\C$. The covariant differentiation of transformation relations $g_{\hal \hbe} = g_{\mu \nu} x^{\mu}_{,\hal} x^{\nu}_{,\hbe}$ along $\C$ gives:
\be \label{eq:transcov} \bar{g}_{\mu \nu} \left( \dfrac{\text{D} e^\mu_{\hj}}{\text{D} s} e^\nu_{\hk} + \dfrac{\text{D} e^\mu_{\hk}}{\text{D} s} e^\nu_{\hj} \right) = 0 . \ee
Simplifying equation~\eqref{e:ref:int2_ordre1} with \eqref{e:ref:epoint} and \eqref{eq:transcov} we obtain:
\be \nonumber \bar{g}_{\hat{0} \hmm , \hj} + \bar{g}_{\hat{0} \hj , \hmm} = \bar{g}_{\alpha \sigma} \bar{\Gamma}\indices{^{\sigma} _{\mu \nu}} u^\alpha e^\nu_{\hj} e^\mu_{\hmm} + \bar{g}_{\alpha \sigma} f\indices{^{\sigma} _{\hj \hmm}} u^\alpha . \ee
Then, using the assertion~\eqref{e:ass2} we obtain:
\be \bar{g}_{\alpha \sigma} \bar{\Gamma}\indices{^{\sigma} _{\mu \nu}} u^\alpha e^\nu_{\hj} e^\mu_{\hmm} + \bar{g}_{\alpha \sigma} f\indices{^{\sigma} _{\hj \hmm}} u^\alpha = 0 . \label{e:ass4} \ee
We define $f_{\hmu \hj \hk} = \bar{g}_{\alpha \sigma} e^\alpha_{\hmu} f\indices{^\sigma _{\hj \hk}}$; one can notice that it is symmetric on the last two indices. Then~(\ref{e:ass3}) and~(\ref{e:ass4}) lead to:
\bea
f_{\hl \hj \hmm} & = - \bar{g}_{\theta \beta} \bar{\Gamma}\indices{^\theta _{\nu \alpha}} e^\beta_{\hl} e^\nu_{\hmm} e^\alpha_{\hj} \nonumber , \\
f_{\hat{0} \hj \hmm} & = - \bar{g}_{\theta \beta} \bar{\Gamma}\indices{^\theta _{\nu \alpha}} u^\beta e^\nu_{\hj} e^\alpha_{\hmm} \nonumber .
\eea
From these two relations we deduce the final form of the second order coefficients~\eqref{e:ref:coeff3}.

\begin{acknowledgements}
The authors are grateful to the referee for his constructive remarks and corrections.
\end{acknowledgements}



%
%

\end{document}

%% file: macro.tex
\newcommand{\be}{\begin{equation}}
\newcommand{\ee}{\end{equation}}
\newcommand{\bea}{\begin{eqnarray}}
\newcommand{\eea}{\end{eqnarray}}

\newcommand{\w}[1]{\bm{#1}}
\newcommand{\dd}{\text{d}} 
\newcommand{\A}{\mathcal{A}}
\newcommand{\M}{\mathcal{M}}
\newcommand{\T}{\mathcal{T}}
\newcommand{\U}{\mathcal{U}}
\newcommand{\Y}{\mathcal{Y}}
\newcommand{\Ol}{\mathcal{O}}

\newcommand{\C}{\mathcal{C}}

\newcommand{\hj}{\hat\jmath}
\newcommand{\hi}{\hat\imath}

\newcommand{\hal}{\hat\alpha}
\newcommand{\hbe}{\hat\beta}
\newcommand{\hk}{\hat{k}}
\newcommand{\hl}{\hat{l}}
\newcommand{\hmm}{\hat{m}}

\newcommand{\hmu}{\hat{\mu}}

